\documentclass[english,aps,prl,twocolumn,showpacs,groupaddress,floatfix,graphics,graphicx]{revtex4-1}
\usepackage[latin9]{inputenc}
\usepackage{amsmath}
\usepackage{amssymb}
\usepackage{graphicx}
\usepackage{verbatim}
\usepackage{braket}
\usepackage{soul}
\usepackage{gensymb}
\usepackage[usenames]{color}

\renewcommand\vec{\boldsymbol}

\begin{document}

\title{Pseudospin Paramagnons and the Superconducting Dome \\ in Magic Angle Twisted Bilayer Graphene
%
% I was hoping that the above title also captures superconductivity suppression by valley paramagnon.
%
%Sublattice-Pseudospin-Paramagnon Mediated Superconductivity \\ in Magic Angle Twisted Bilayer Graphene
} 

\author{Chunli Huang, Nemin Wei, Wei Qin, and Allan~H.~MacDonald}
\affiliation{Department of Physics, University of Texas at Austin, Austin TX 78712}

\date{\today} 

\begin{abstract}
We present a theory of superconductivity in magic-angle twisted bilayer graphene in which attraction is generated 
between electrons on the same honeycomb sublattice when the system is close to a sublattice polarization instability. 
The resulting Cooper pairs are spin-polarized valley-singlets.
Because the sublattice polarizability is mainly contributed by interband fluctuations, superconductivity occurs over a wide range of  moir\'e filling fraction. It is suppressed by i) applying a sublattice polarizing field (generated by an
aligned BN substrate) or ii) changing moir\'e band filling to favor valley polarization. The enhanced intrasublattice attraction close to sublattice 
polarization instability is analogous to
enhanced like-spin attraction in liquid $^3$He near the melting curve and the enhanced valley-singlet repulsion close to valley-polarization instabilities is analogous to enhanced spin-singlet repulsion in metals that are close to a ferromagnetic instability.
We comment on the relationship between our
pseudospin paramagnon model and the rich phenomenology of superconductivity in twisted bilayer and multilayer 
graphene. 
\end{abstract}

\maketitle

\begin{figure}
    \centering
    \includegraphics[width=1\columnwidth]{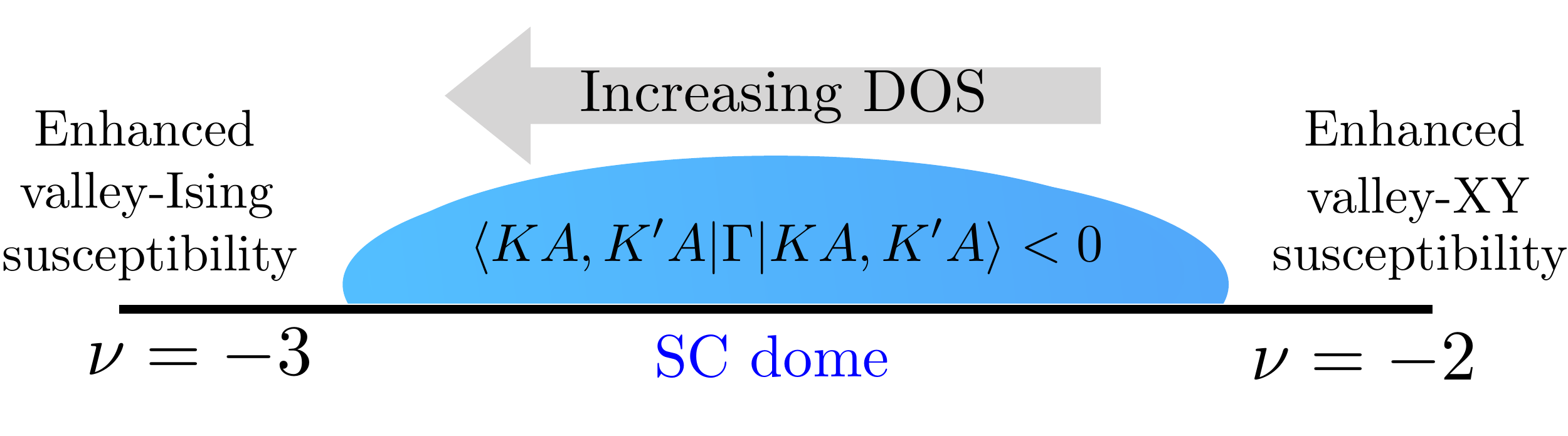}
    \caption{Exchange-enhanced sublattice pseudospin polarizatibility leads to an attractive intrasublattice
    interaction that can support spin-triplet, valley-singlet superconductors across a wide range of moir\'e band filling fraction $\nu$. 
    Superconductivity is suppressed when $\nu$ is tuned close to valley polarization instabilities and when the density-of-states (DOS) is small.}
    \label{fig:main}
\end{figure}

\textit{Introduction:--}
Possible explanations for superconductivity in magic-angle twisted bilayer (trilayer) graphene 
MATB(T)G  \cite{cao2018unconventional,park2021tunable} are increasingly constrained by experimental data \cite{yankowitz2019tuning,lu2019superconductors,saito2020independent,polshyn2019large,
cao2020strange,de2021gate,liu2021tuning,wong2020cascade,stepanov2020competing,zondiner2020cascade,saito2021isospin,cao2021pauli,choi2021interaction,choi2021correlation,wong2020cascade}.
The normal state properties revealed by the weak-field Hall effect and 
magnetoresistance oscillations \cite{saito2021isospin} 
are particularly telling; the strongest superconductivity occurs within the moir\'e band filling
interval $\nu \in (-3,-2)$, within which the system has a hole-like Fermi surfaces surrounding $|\nu|-2$ states 
per moir\'e period.  The implied Fermi surface reconstruction points to symmetry breaking that depopulates two
of four spin/valley flavors, an interpretation that is reinforced by 
Landau fans that are only doubly degenerate for $\nu \in (-3,-2)$ in spite of the systems four-fold spin/valley band 
degeneracy.  The normal state symmetry breaking provides a natural explanation for
large in-plane critical magnetic fields \cite{cao2021pauli} that are $ \sim 2-3$ times larger than
the Clogston-Chandrashekar limit of spin-singlet superconductors.  As summarized in Fig.~\ref{fig:main}, superconductivity is suppressed as $\nu$ approaches
$-3$ and as $\nu$ approaches $-2$, filling factors at which
the system is known to tend toward valley-ordered states,
forming a dome in the system's $(T,\nu)$ phase diagram that is reminiscent of those observed in cuprate superconductors. 
For example, the superconducting critical temperature $T_c$ of device 1 in Ref.~\cite{saito2020independent} 
is peaked at $\nu\sim-2.4$, and vanishes near $\nu\rightarrow-2.7$ on the low-filling-factor side and 
near $\nu\rightarrow-2.2$ on the high filling factor side.  The suppression of superconductivity on the low-filling factor 
side occurs in spite of an increasing Fermi level density-of-states,
and therefore argues against a mechanism, like phonon-dressing, in which the pairing glue is external to the electron system.  
Superconductivity actually seems to be suppressed when the Fermi level is close to a the flat valence band 
van Hove singularity energy \cite{stepanov2020competing,park2021tunable}.  These experimental observations 
point to an electronic pairing mechanism.  
%
%
%around $-3<\nu<-2$ and it has a critical temperature $T_c$ that exhibits an interesting hole-density dependence similar to the superconductivity dome in Cuprate superconductors. Indeed, the critical temperature $T_c$ is peaked at a device-dependent optimal hole-density between $-3<\nu<-2$ and it decreases when more holes are added ($\nu\rightarrow-3$) or removed ($\nu\rightarrow-2$), as shown schematically in Fig.~\ref{fig:main}. 
%Note most of the superconductivity domes are skewed towards the $\nu=-2$ side and some devices have additional smaller domes on the $-2<\nu<-1$ side \cite{saito2020independent}, take device 1 of Ref.~\cite{saito2020independent} for instance, the superconductivity dome appears at $\nu\sim-2.2$, peaks at $\nu\sim-2.3$ and disappears at $\nu\sim-2.6$. 
%Scanning tunneling spectroscopy \cite{choi2021correlation} found that the moire flat bands become highly $\nu$ dependent because of the mean-field Hartree-Fock effect.

%Suppression of $T_c$ with large DOS suggests the dominant pairing mechanism is not via the exchange of virtual phonon because phonon dispersion is mostly determined by the properties of ions (plus some screening coming from the electrons) so it would predict a $T_c$ that  more or less follows the DOS. Moreover, superconductivity is either not observed at the van-Hove singularity \cite{stepanov2020competing} or its $T_c$ is suppressed by proximity to the van-Hove singularity \cite{park2021tunable}. These experimental observations are difficult to interpret with phonon-mediated superconductivity.

In this Letter, we argue that the superconducting properties of MATBG are consistent with 
a sublattice-pseudospin paramagnon pairing mechanism, and interpret the shape of the superconducting dome 
in terms of sublattice and valley paramagnons.
%These pseudospin fluctuations are strong because of exchange-effect and they enhance (suppress) spin-triplet (singlet) pairing hence, the resulting superconductivity emerges from the normal state where the spin from opposite valley polarized in the same direction.
Our main message is summarized in Fig.~\ref{fig:main}.
%The attractive interaction is generated by the pseudospin fluctuations that are strongly exchange-enhanced in a spin-ferromagnetic normal state, so they naturally to a spin-triplet superconductor and thus consistent with the observed Pauli-limit violation. 

\begin{figure}
    \centering
    \includegraphics[width=1\columnwidth]{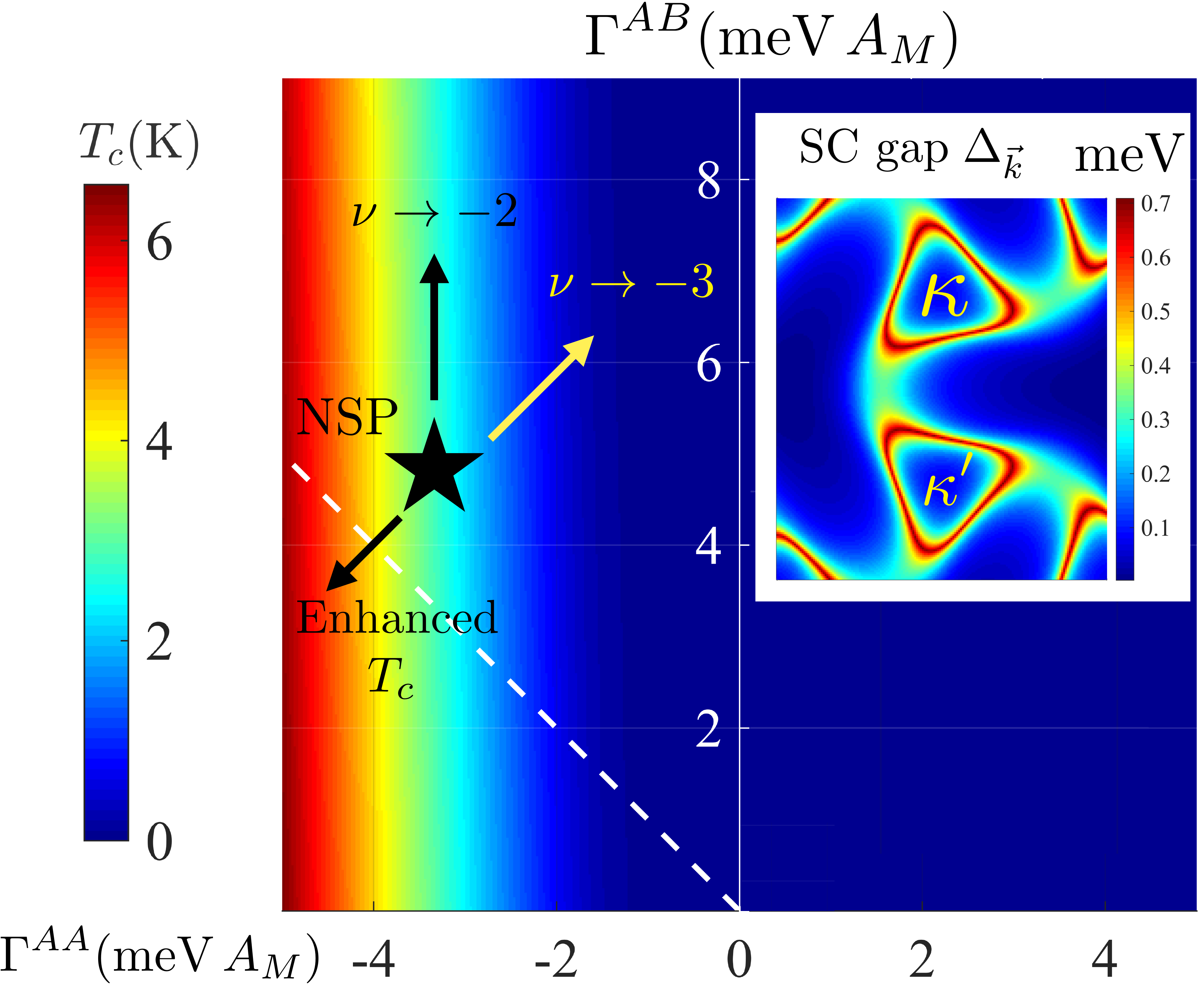}
    \caption{
    $T_c$ {\it v.s.} $\Gamma^{AA}$ and $\Gamma^{AB}$. $A^{a} =(\Gamma^{AA}-\Gamma^{AB})/2$ is attractive and strong 
    in nearly sublattice polarized (NSP) metals, whereas $A^s=(\Gamma^{AA}+\Gamma^{AB})/2$ is repulsive, but weak if screening is 
    strong. Superconductivity occurs at low temperature when $\Gamma^{AA}$ is attractive, even though $\Gamma^{AB}$ is repulsive.  The physically accessible parameter range above the dashed white $A^s=0$ line includes superconductivity states.
    The inset plots the pairing self-energy $\Delta_{\vec{k}}$ {\it vs.} $\vec{k}$ in the moir\'e Brillouin-zone for 
    a continuum model with $t_{AA}/t_{AB}=0.7$, at $\theta=1.1^{\circ}$, at $\nu=-2.4$ \cite{SM}.  $\Delta_{\vec{k}}$ is largest near the normal 
    state Fermi surfaces centered on $\kappa$, $\kappa'$.
    When the system is close to a valley-Ising instability (near $-3$) or when it is sublattice polarized (near $\nu=+3$)  \cite{serlin2020intrinsic,sharpe2019emergent}, $\nu=+1$ \cite{stepanov2020competing}), $\Gamma^{AA}$ will increase 
    and superconductivity will weaken. When $A^s$ is decreased by the acoustic-phonon mediated interaction and/or screening of the 
    long-range Coulomb interactions, $T_c$ is enhanced. These trends are indicated by arrows and discussed in the main text.
    %For $\Gamma^{AB}>0$, superconductivity is observed only when $\Gamma^{AA}<0$ and is strongest close to the sublattice ferromagnetic instability where $-\Gamma^{AA} \gg \Gamma^{AB}>0$. When the filling fraction $\nu$ is tuned close to the valley-Ising (XY) instability near $\nu=-3$ ($\nu=-2$), the attraction $|\Gamma^{AA}|$ weakens.
    }
    \label{fig:tcdiagram}
\end{figure}

\textit{Pseudospin-paramagnons (PPM) and fermion pairing:-}
Our theory of MATBG superconductivity is guided by experiments \cite{cao2021pauli,saito2021isospin} 
and inspired by what is known about the relationship between fermion pairing in liquid $^3$He 
and the strongly enhanced paramagnetic spin susceptibilities that appear near the solidification curve~\cite{levin1983phenomenological}.
The enhanced susceptibility leads to a low-frequency neutral excitation spectrum that is dominated by 
\textit{paramagon} peaks, damped collective modes that generate a strong attraction between like-spin nuclei, and leads to 
anisotropic spin-triplet Anderson-Brinkman-Morel \cite{anderson1973anisotropic} superfluidity.
Spin-fluctuation mediated superfluidity in $^3$He is succinctly captured by the paramagnon model,
which uses a single exchange-interaction parameter $I$ to describe the enhanced spin susceptibility \cite{berk1966effect}. 
The paramagnon model accounts quantitatively for the $T^3\ln T$ contribution to heat capacity in $^3$He \cite{doniach1966low},
and has been extended to understand the interplay between magnetism and superconductivity in metals,
including heavy-fermion materials \cite{anderson1984heavy,moriya2006developments,RevModPhys.84.1383}.

Unlike nuclei in $^3$He, which possess only a spin degree of freedom, flat-band electrons in 
MATBG possess a spin(s)$\times$valley($\tau$)$\times$sublattice($\sigma$) pseudospin octet 
that generates eight distinct states for each momentum and 
an abundance of potential pairing channels that has to be winnowed.  Some progress can be achieved by taking note 
of the time-reversal symmetry property $\epsilon_{K}(\vec{k})=\epsilon_{K'}(-\vec{k})$, 
which strongly favors Cooper pairing between electrons in opposite valleys $K$ and $K'$ ($\epsilon_{K}(\vec{k})\ne \epsilon_{K}(-\vec{k})$).
If we assume \cite{cao2021pauli,saito2021isospin}  that the normal 
state for $\nu < -2$ is ferromagnetic (in spin), we can conclude that the Cooper pairs must be spin-polarized valley-singlets. That still leaves the sublattice pseudospin, and the need to overcome the strong sublattice independent 
Coulomb repulsion. A route to superconductivity is provided by the properties of the flatband spinors which imply, as illustrated in Fig.~2, 
that superconductivity occurs when the intrasublattice interaction is attractive ($\Gamma^{AA}<0$)
even if the intersublattice interaction is strongly repulsive ($\Gamma^{AB}>0$) \footnote{It is also true that 
attractive intersublattice interactions $\Gamma^{AB}<0$ produce superconductivity when the intrasublattice interaction 
is repulsive.  See \cite{SM} for a discussion of this case}. 
This property of magic-angle superconductivity is analogous to the robustness of spin-triplet 
superconductivity against repulsive opposite-spin interactions in systems with weak spin-orbit coupling \cite{anderson2018theory},
and can be traced to the $D_6$ point-group symmetry of the band Hamiltonian which decouples like-sublattice and unlike-sublattice pairing in the linearized gap equation \cite{SM}. 
We show below that intrasublattice attraction is generated
when the system is close to a spontaneous sublattice polarization instability.
This instability is common in graphene multilayers because sublattice polarization simplifies 
the spinor content of occupied states and increases exchange energies \cite{macdonald2012pseudospin, velasco2012transport}.
The same analysis that shows that sublattice PPM bolster superconductivity, shows that valley PPM are obstructive.

%\textbf{CL to Allan: I modified your paragraph in red to blue. Please see if you agree.}
%\textcolor{red}{
%A route to superconductivity is provided by the properties of flat band pseudospinors which 
%imply, as illustrated in Fig.~\ref{fig:tcdiagram}, that superconductivity occurs if either the interaction between electrons on the same sublattice or the interaction between electrons on opposite sublattices is attractive.  There is no need for both interactions to be attractive, or even for the sum of the two interactions to be attractive. This property, can be traced to the 
%$\mathcal{C}_2$ inversion symmetry of the band Hamiltonians which decouples pairing on the same sublattice and   pairing on opposite sublattices, as illustred in Fig.~\ref{fig:tcdiagram}. We show below that attractive interactions between electrons on like sublattices are generated
%when the system is close to a spontaneous sublattice polarization instability. These instabilities are common in graphene multilayers because sublattice polarization simplifies 
%the spinor content of occupied states and increases exchange energies. \cite{macdonald2012pseudospin}
%The same analysis that shows that sublattice PPMs bolster superconductivity, shows that valley PPMs are obstructive.  }

\textit{Sublattice-Dependent Effective Interactions:-}. The reduced pairing Hamiltonian for spin $\uparrow$ 
electrons in opposite valleys is 
\begin{align} \label{eq:H_red}
H_{\text{red}}=&\sum_{\vec{k}\vec{k}',\sigma_i}\; \bigg[\Gamma_{\vec{k}\vec{k}'}(\sigma_{1}\sigma_{4};\sigma_{2}\sigma_{3}) \nonumber \\
&\times c_{\vec{K}+\vec{k}'\sigma_{1}\uparrow}^{\dagger}c_{-\vec{K}-\vec{k}'\sigma_{2} \uparrow}^{\dagger}c_{-\vec{K}-\vec{k}\sigma_{3}\uparrow}c_{\vec{K}+\vec{k}\sigma_{4}\uparrow} \bigg],
\end{align}
where $\Gamma$ is the particle-particle channel irreducible 4-point vertex function estimated below \cite{vignale1985effective,bickers2004self}, 
$\vec{k}$ labels a state in the moir\'e valence band, and $\sigma = (A,B)$ labels sublattice.
As we explain below, the PPM contributions to $\Gamma$ are dominantly diagonal in
sublattice at each vertex.  This property motivates a model in which the dependence of $\Gamma$ on $\vec{k},\vec{k}'$ is neglected 
and $\mathcal{C}_2$ symmetry is recognized: 
\begin{equation}  \label{eq:Gamma_4by4}
\Gamma_{\vec{k}\vec{k}'}(\sigma_{1}\sigma_{4};\sigma_{2}\sigma_{3}) \to \delta_{\sigma_{1},\sigma_{4}} \delta_{\sigma_{2},\sigma_{3}}
\big[ \Gamma^{AA} \delta_{\sigma_{1},\sigma_{2}} + \Gamma^{AB} \delta_{\sigma_{1},-\sigma_{2}}\big].
\end{equation}
As illustrated in Fig.~\ref{fig:tcdiagram}, superconductivity occurs in this model when the same 
sublattice effective interaction $\Gamma^{AA}$ is attractive.
The superconductivity gap is approximately uniform on the valley-projected FS (centered around $\kappa$ and $\kappa'$), as shown in the inset. 
$T_c$ is not suppressed by repulsive  $\Gamma^{AB}$ interactions because the susceptibility for pairs on opposite sublattices 
averages to zero on the Fermi surface. Motivated by this observation, we now address the relationship of $\Gamma^{AA}$ to PPMs.
%The resulting superconducting gap is identical in both 
%The superconducting gap is intra-sublattice and s-wave,  although inter-sublattice components of the Cooper pair wave functions are notably in the higher angular momentum channels and help avoid the short range inter-sublattice repulsion.}\textcolor{blue}{More explanations here?}

\begin{figure*}
    \centering
    \includegraphics[width=2.0\columnwidth]{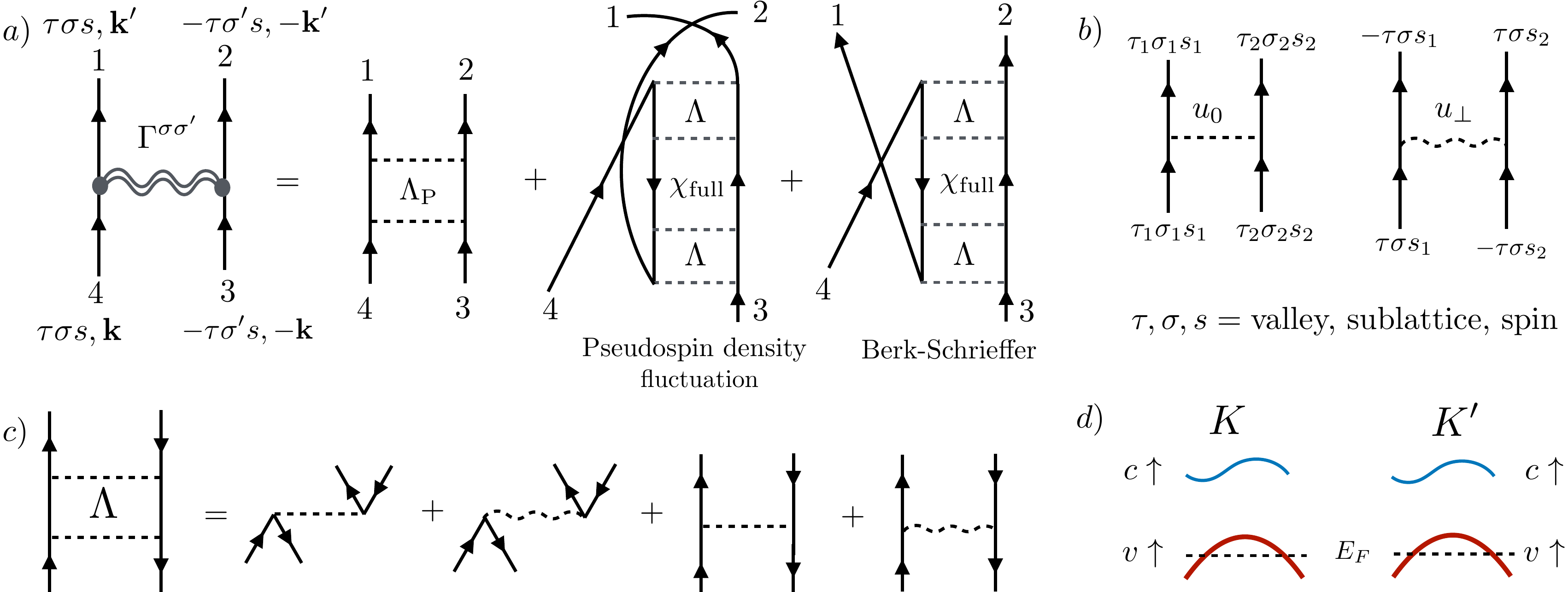}
    \caption{a) The interaction potential $\Gamma$ relevant to superconductivity has three contributions \cite{vignale1985effective,bickers2004self}: a particle-particle vertex function $\Lambda_{\text{P}}$, a double-crossing diagram and the Berk-Schrieffer (single-crossing) diagram. The crossing diagrams relate $\Gamma$ to paramagnon interactions that diverage at particle-hole instabilities. 
    %For a Cooper pair with zero center-of-mass momentum, the external legs are $(1;2;3;4)=(\vec{K}+\vec{k}',\sigma_1,\uparrow; -\vec{K}-\vec{k}',\sigma_2,\uparrow;-\vec{K}-\vec{k},\sigma_3,\uparrow ;\vec{K}+\vec{k},\sigma_4,\uparrow)$ where $\sigma_i$ is a sublattice label. 
    b) $u_0>0$ is a pseudospin-independent density-density interaction and  $u_\perp>0$ is an inter-valley scattering potential. 
    c) The particle-hole vertex function $\Lambda$ consists of the direct and exchange scattering terms from both $u_{0}$ and $u_\perp$. d) Schematic Fermi surface for $-3<\nu<-2$ where $c(v)$ are the spin and valley projected conduction (valence) band. }
    \label{fig:diagrams}
\end{figure*}

Recognizing that the computation of $\Gamma$ from first-principles is a 
formidable challenge even in relatively simple systems \cite{anderson2018theory,nava2012equation}, 
we follow a phenomenological approach similar in spirit to that employed successfully in $^3$He \cite{anderson1984heavy}.
In that case spin-rotational invariance allows the spin-dependent effective interaction 
to be parameterized by just two scattering 
amplitudes, one each for the spin-symmetric and spin antisymmetric channel,
(\textit{i.e.~}$\Gamma =A^s+A^a\vec{s_1}\cdot \vec{s_2}$ ~\cite{levin1983phenomenological}).
In  MATBG, the Hamiltonian is not invariant under rotations of the sublattice pseudospins.
When the system is close to a sublattice-polarization instability, however, 
effective interactions that are diagonal in sublattice dominate and the analogous expressions are 
\begin{equation}
    \Gamma^{AA}=A^{s}+A^{a} \;\; , \;\; \Gamma^{AB}=A^{s}-A^{a}.
\end{equation}
We estimate $A^{s}$ and $A^{a}$ by assuming a simple $\delta$-function interaction model 
in order to sum the Feynman diagrams specified by Fig.~\ref{fig:diagrams}. The irreducible vertex function $\Lambda$ accounts for both direct and exchange scattering  ({\it i.e.~}generalized-random-phase approximation) generated by $u_{0}$ and $u_{\perp}$.
$u_0$ is the fully symmetric Coulomb interaction while the 
valley-exchange scattering term $u_{\perp}>0$ breaks independent 
spin-rotation symmetry in opposite valleys and selects the spin-ferromagnet over other possibilities as the flavor-polarized normal state. \footnote{We use crossing-symmetry \cite{bickers2004self} to relate the vertex functions in the particle-particle channel ($\Lambda_P$) to the one in particle-hole channel ($\Lambda$).}

Ignoring for the moment the Berk-Schrieffer transverse fluctuation diagram
in Fig.~\ref{fig:diagrams}, we find that 
\begin{align}
    A^{s} = \frac{1}{2} \left( T^{\tau_0\sigma_0}-T^{\tau_z\sigma_0} \right)\;,\;
    A^{a} = \frac{1}{2} \left( T^{\tau_0\sigma_z}-T^{\tau_z\sigma_z} \right),
\end{align}
where 
%{\bf Allan:  I don't know why we specify $\omega=0$ here.  Whey don't we just write $\omega$ instead of $\omega=0$.  The small $\omega$ expansion is discussed later.}\textcolor{blue}{Nemin: Rigorously speaking, the decoupling of $\tau_0\sigma_0$ and $\tau_0\sigma_z$ channels is valid only in two limits: 1) $\vec{q}=0$ because of $C_{2}$ symmetry 2) $\omega=0$ because $C_2T$ requires $\chi^{\tau_0\sigma_0,\tau_0\sigma_z}_{\vec{q},\omega=0}=-\chi^{\tau_0\sigma_0,\tau_0\sigma_z}_{\vec{q},-\omega}$. Since in the numeric calculation we will use a $\vec{q}-$independent interaction, I think the second limit is more relevant to our discussion so we have to constrain ourselves to $\omega=0$.}
\begin{eqnarray}
T^{\tau_0\sigma_0} &=&  \frac{1}{2}\frac{3u_{0}-u_{\perp}}{1+(3u_{0}-u_{\perp})\chi_{\vec{k}-\vec{k'},\omega}^{K\sigma_{0},K\sigma_{0}}}, \label{eq:tau0sigma0} \\
T^{\tau_z\sigma_0} &=& -\frac{1}{2}\frac{u_{0}-u_{\perp}}{1-(u_{0}-u_{\perp})\chi_{\vec{k}-\vec{k'},\omega}^{K\sigma_{0},K\sigma_{0}}}, \label{eq:tauzsigma0} \\
T^{\tau_0\sigma_z} &=& -\frac{1}{2}\frac{u_{0}+u_{\perp}}{1-(u_{0}+u_{\perp})\chi_{\vec{k}-\vec{k'},\omega}^{K\sigma_{z},K\sigma_{z}}},  \label{eq:tau0sigmaz} \\
T^{\tau_z\sigma_z}  &=& -\frac{1}{2}\frac{u_{0}-u_{\perp}}{1-(u_{0}-u_{\perp})\chi_{\vec{k}-\vec{k'},\omega}^{K\sigma_{z},K\sigma_{z}}}, \label{eq:tauzsigmaz}
\label{eq:tauzsigmaz}  
\end{eqnarray}
Here $\chi_{\vec{q},\omega=0}^{O,O^\dagger}$ is a bare susceptibility for pseudospin polarization $O$. We discuss the valley-diagonal, sublattice even ($\chi_{\vec{q},\omega=0}^{K\sigma_{0},K\sigma_{0}}$) and 
sublattice odd ($\chi_{\vec{q},\omega=0}^{K\sigma_{z},K\sigma_{z}}$) susceptibilities further below.

%At lowest order Eq.~\eqref{eq:tau0sigma0}--\eqref{eq:tauzsigmaz} account for the direct and exchange scatterings associated with $u_0$ and $u_{\perp}$.  

The random phase sums decouple contributions that are proportional to different products of valley and sublattice Pauli matrices ($\tau_0$ or $\tau_z$ and $\sigma_0$ or $\sigma_z$) of the interacting particle-hole pairs and yield pseudospin-dependent interactions that are sums of the four different geometric series. 
In each of the $T^{\tau_a\sigma_b}$'s a bare interaction is modified by a dressing factor (\textit{i.e.~}denominator) identical to the one that modifies the corresponding susceptibility.  For the valley and sublattice independent interaction 
$T^{\tau_0\sigma_0}$ the bare interaction is dominated by the bare repulsive Coulomb interaction $u_0$ and the modification is suppression
due to static screening.  For all other interaction channels, the dominant bare contribution is  
an attractive exchange contribution and the modification factor yields enhancement.  At lowest order $\Gamma^{AA}=u_0-u_{\perp} $ 
and $\Gamma^{AB}=u_0$ are both repulsive.  Superconductivity is possible when $T^{\tau_0\sigma_z}$, which contributes attractively to 
$\Gamma^{AA}$, has much larger enhancement factors than $T^{\tau_z\sigma_0}$ and $T^{\tau_z\sigma_z}$, both of which contribute repulsively and
diverge at valley polarization instabilities.  We conclude that attractive intra-sublattice effective interactions, and hence superconductivity, is likely
when the system is close to a valley-independent sublattice polarization instability but far from a valley-polarization 
instabilities.

We have so far neglected the Berk-Schrieffer (BS) transverse valley fluctuations 
indicated in Fig.~\ref{fig:diagrams}. 
The particle-hole channels in the BS diagrams
decouple into four channels $\tau_{+}\sigma_{0,x,y,z}$ in the long wavelength limit. They change
\begin{equation}
    A^{s,a}\rightarrow A^{s,a} \mp \frac{1}{2} \sum_{i=x,y}T^{\tau_+\sigma_i} -\frac{1}{2} \sum_{\delta=0,z}T^{\tau_+\sigma_\delta}.
\end{equation}
Because the susceptibilities in the $\delta = 0,z$ channels excite particle-hole pair from opposite Chern bands, their average transition matrix elements are small \cite{bultinck2020ground, lian2021iv}. 
As shown in  Ref.~\cite{SM}, they are in fact smaller than the susceptibility in the pseudospin density fluctuation channels.
%Because the susceptibilities in the $\delta = 0,z$ channels are related to the overlap of the Bloch wavefunctions from opposite Chern bands \cite{bultinck2020ground, lian2021iv}, in the pseudospin density fluctuation channels \cite{SM}.
%
The other two channels in the BS diagrams ($i=x,y$) are off-diagonal in sublattice and therefore do not affect the attractive interaction $\Gamma^{AA}$. As a result, the net effect of transverse valley fluctuations (which can be strong close to intervalley coherent order \cite{bultinck2020mechanism, zhang2019nearly, lee2019theory} ) is to increase $\Gamma^{AB}$ in the vertical direction shown in Fig.~\ref{fig:tcdiagram} and do not modify the $T_c$. Hence, transverse valley fluctuations can be neglected in our analysis.
In making these estimates of importance, we are not considering the possibility of a nodal superconducting gap generated by repulsive interactions with strong momentum dependence - something that cannot be ruled out.
%so they can be neglected as $T_c$ is largely independent of $\Gamma^{AB}$ , \textit{c.f.~} Fig.~2.
%We are not considering the possibility of a nodal superconducting gap generated by repulsive interaction with strong momentum dependence - something that cannot be ruled out.

In the analysis of spin-fluctuation mediated Cooper pairing in liquid $^3$He and other Fermi-systems with weak spin-orbit coupling, there is often no need to distinguish the longitudinal spin-fluctuation double-crossing diagram from the transverse spin-fluctuation Berk-Schrieffer diagram because the pairing interaction is invariant under SU$(2)$ spin rotation.
This distinction is important in our present analysis because the pseudospin ($\tau \otimes \sigma$) anisotropic energy arose from the band Hamiltonian is very large and as a result, the transverse and longitudinal pseudospin fluctuations can contribute independently to the pairing interaction. 

\textit{The Superconducting Dome:--}. We are now in a position to
explain how the distinct filling factor dependencies of the different $T^{\tau_{a}\sigma_{b}}$'s forms the superconducting dome. An enhanced sublattice polarization susceptibility, which strengthens an attractive interaction, is present when 
the corresponding Stoner criterion $(u_{0}+u_{\perp}) \chi_{\vec{q},\omega=0}^{K\sigma_{z},K\sigma_{z}}=1$,
is nearly satisfied.  Repulsive interactions are strengthened when 
one of the valley polarization susceptibilities, with Stoner criteria
$(u_{0}-u_{\perp})\chi_{\vec{q},\omega=0}^{K\sigma_{0},K\sigma_{0}}=1$ or $(u_{0}-u_{\perp})\chi_{\vec{q},\omega=0}^{K\sigma_{z},K\sigma_{z}}=1$,
are close to being satisfied.  
Because $u_{0}+u_{\perp}$ is larger than $u_{0}-u_{\perp}$  the attractive $\tau_0\sigma_z$ channel interaction 
is always stronger than the repulsive $\tau_z\sigma_z$ channel.
%; the competition between these two channels is discussed further below.
Physically, the $\tau_0\sigma_z$ susceptibility is enhanced and the $\tau_z\sigma_z$ susceptibility suppressed by 
$u_{\perp}$ because its exchange energy is maximized when opposite valleys have identical sublattice polarization.
The competition between normal and superconducting 
states is therefore between the sublattice polarization Stoner criteria, 
which involves the $\chi_{\vec{q},\omega=0}^{K\sigma_{z},K\sigma_{z}}$ polarization, 
and the valley polarization Stoner criterion, which involves $\chi_{\vec{q},\omega=0}^{K\sigma_{0},K\sigma_{0}}$.  
Because of $\mathcal{C}_2T$ symmetry (where $T$ is a spinless time-reversal symmetry) these two polarizations have very distinct dependencies on band filling.
In particular $\chi_{\vec{q}\rightarrow0,\omega=0}^{K\sigma_{z},K\sigma_{z}}$ does not have an intraband contributions because the 
expectation value of $\sigma_{z}$ is zero in all moir\'e band states. 
Its value is therefore not related to 
peaks in the band density-of-states and, as illustrated in Fig.~3d, should instead 
be larger closer to $\nu=-2$ where the majority spins are 
nearly half-filled and the interband transition phase space is therefore largest.
On the other hand, $\chi_{\vec{q}\rightarrow0,\omega=0}^{K\sigma_{0},K\sigma_{0}}$ is proportional to the Fermi level density of states 
and increases rapidly as $|\nu|-2$ increases and the Fermi level approaches the Van Hove peak in the 
valence moir\'e band.  
We attribute the suppression of superconductivity on the large 
$|\nu|-2$ side of the superconducting dome to the increasingly strong repulsive interaction contribution to $A^{s}$ from $T^{\tau_z\sigma_0}$, which is proportional to the valley polarization susceptibility enhancement factor and suppresses superconductivity, 
as indicated in Fig.~\ref{fig:tcdiagram}. 
%On the other hand, we attribute the suppression of superconductivity on the small $|\nu|-2$ side of the dome to the combined effect of a decreasing Fermi-level density-of-states and Berk-Schrieffer transverse valley fluctuations in proximity to the intervalley-coherent orders \cite{bultinck2020mechanism, zhang2019nearly, lee2019theory}. The latter mainly increases $\Gamma^{AB}$ in the vertical direction shown in Fig.~\ref{fig:tcdiagram} and is not 
On the other hand, we attribute the suppression of superconductivity on the small $|\nu|-2$ side of the dome to the decreasing Fermi-level density-of-states.

We emphasize that our picture of superconductivity in graphene moir\'e superlattice 
requires that the normal state is nearly sublattice polarized metal over a wide range of filling fraction.
Enhanced sublattice polarization, and sometimes sublattice polarization instabilities, are in fact common 
in all graphene multilayer electron gas systems \cite{macdonald2012pseudospin}.
 For example, neutral Bernal bilayer graphene has $\sigma_z s_z$ order, \textit{i.e.~} spin-dependent sublattice-polarization in the antiferromagnetic state~\cite{velasco2012transport}.
In twisted bilayers the gaps between flat and remote bands open an opportunity for flavor polarization and hence for 
sublattice polarization not only near neutrality but also near integer $\nu$.
In moir\'e flat bands itinerant exchange energies favor maximum spin-polarization before orbital-polarization, 
allowing the majority spin projected conduction and valence bands to mix and
produce finite sublattice polarization over a wide range of filling fraction.

\textit{Summary and Discussions:--} Paramagnon-mediated pairing 
physics in MATBG is enriched by the relevance of spin, valley, and sublattice 
pseudospins. Because the valley-projected bands are not time-reversal invariant, 
pairing is likely to occur between time-reversal partner states in opposite valleys and 
yield valley-singlets. 
We have shown that valley-singlet superconductivity occurs in MATBG when the 
effective interaction between electrons on the same sublattice is attractive, and that this condition is satisfied 
when the system is close to a sublattice-polarization instability, but far from a valley-polarization instability.
There is no need for the interaction between electrons on opposite sublattice, or the total interaction summed over sublattices, to be attractive. 
%For paired electrons that are drawn from opposite valleys, the attractive intrasublattice interactions close to sublattice polarization instabilities and the repulsive interactions close to valley-polarization instabilities are respectively analogous to spin-triplet attraction \cite{anderson1973anisotropic} and spin-singlet repulsion \cite{berk1966effect}  close to a ferromagnetic instability.
%
Enhanced intrasublattice attraction close to sublattice 
polarization instabilities is analogous to the
enhanced like-spin attraction in $^3$He \cite{anderson1973anisotropic} near the melting curve, 
and enhanced valley-singlet repulsion close to valley-polarization instabilities
is analogous to enhanced spin-singlet repulsion \cite{berk1966effect} in metals that are close to a ferromagnetic instability.
Together these two effects explain the prominent superconducting domes 
seen in most MATBG samples inside the filling factor interval $\nu \in (2,3)$ \footnote{When the filling factor $\nu$ approaches $\nu=-3$, the enhanced valley polarizability not only generates a valley-singlet repulsion but also an attraction between electrons in the same valley. When this attraction leads to condensation of Cooper pairs from the same valley, the resulting superconducting order parameter cannot be uniform-s-wave because of Fermi statistics. (We assume the normal state is always spin-polarized). This means the order parameter might undergo a transition from valley-singlet to valley-triplet as $\nu\rightarrow -3$ and might explain the $U-V$ transition recently observed in local-tunneling spectroscopy \cite{kim2021spectroscopic}. In this scenario, valley-triplet superconductivity close to $\nu=-3$ is eventually suppressed by the enhanced pair-breaking (trigonal-warping) potential from the band-Hamiltonian.}.

We now comment on other aspects of the phenomenlogy of MATBG superconductivity seen 
through the pseudospin paramagnon lens:
  
\noindent 
i) \textit{Influence of Boron-Nitride (hBN) Alignment}:-- Aligned hBN induces 
a finite sublattice polarization in a graphene sheet, and weakens sublattice pseudospin fluctuations.
Since all other contributions to the intrasublattice interaction are repulsive, this theory 
predicts that superconductivity is suppressed by hBN alignment, in agreement with current 
experimental findings.

\noindent
ii)  \textit{Coulomb screening}:-- When MATBG devices are surrounded by nearby material that is 
conducting \cite{saito2020independent,stepanov2020untying,liu2021tuning} or has a large dielectric constant
 \cite{arora2020superconductivity}, insulating states at integer filling factors become less prominent in the 
 phase diagram and superconductivity is found over a broader range of filling factors.  Insulating states 
 are understood in terms of exchange-splitting between bands associated with different flavor states that is larger
 than the flatband bandwidth.  Since the exchange splittings are dominated by the Coulomb interaction $u_0$, they are  
 expected to be reduced by enhanced environmental screening.  The reduction in insulating state gaps seen 
 experimentally is therefore expected.  On the other hand, the net attractive interaction contributed by 
$T^{\tau_0\sigma_z}- T^{\tau_z\sigma_z}$ scales with the intervalley-exchange interaction $u_{\perp}$, which
should not be influenced by environmental screening.  
In addition the enhancement factors in $T^{\tau_0\sigma_z}$ and $T^{\tau_z\sigma_z}$, which together contribute attractively to $\Gamma^{AA}$, 
both involve the $\chi_{\vec{q},\omega}^{K\sigma_{z},K\sigma_{z}}$ susceptibility which does not have a $\vec{q}=0$ Fermi surface contribution.
$\chi_{\vec{q},\omega}^{K\sigma_{z},K\sigma_{z}}$ is instead dominated by interband fluctuations
that are less sensitive to remote screening.
On the other hand, $T^{\tau_0\sigma_0}$, the repulsive screened Coulomb interaction, 
is certainly weakened by enhanced dielectric screening. The net result of enhanced remote screening could
therefore be to enhance superconductivity,
as sometimes observed experimentally. 
Although both superconductivity and insulating behavior require interactions,
remote screening influences only $u_0$, and can therefore 
have opposite influences on the two states.
 
 \noindent
 iii) \textit{Superconductivity near $\nu=0$}:-- Although superconductivity is most robustly observed for $\nu \in (2,3)$, the hole-like and electron-like Fermi surfaces 
 on opposite side of $\nu=0$ also sometimes host superconductivity as shown in Ref.~\cite{lu2019superconductors}. 
 Since our proposed theory of superconductivity between $-3<\nu<-2$ only involves states in a spin-projected Hilbert space (\textit{c.f.}~Fig.~3d), it can be extended to $-1<\nu<+1$ by assuming an identical pairing mechanism occurs in in both
 spin projected Hilbert spaces.  This state would be similar to the equal spin pairing state in $^3$He.
 The stronger superconductivity for $-3<\nu<-2$, could be attributed to band-renormalizations 
 that flatten bands away from half-filling. \cite{choi2021interaction,PhysRevB.100.205113,guinea2018electrostatic,xie2020weak}. 

\noindent
iv) \textit{Particle-hole asymmetry of superconductivity}:-- Experiment shows that the tendency toward valley 
polarization, which opposes superconductivity, is stronger for electron-doped than for hole-doped 
MATBG.  This tendency is due in part to conduction bands that are less dispersive than the valence bands, and have a 
different shape \cite{choi2021interaction,PhysRevB.100.205113,guinea2018electrostatic,xie2020weak}.  
It is quite likely that the normal state on the electron-doped side is a sublattice ferromagnet, as is supported by the observation of the anomalous Hall effect over a wide range of filling factors 
near $\nu=3$ in devices with aligned hBN \cite{serlin2020intrinsic,sharpe2019emergent} 
and in $\nu=1$ \cite{stepanov2020competing} without aligned hBN. 
In addition, the valley PPMs that suppress superconductivity are certain to be stronger on the electron-doped side. 

v) \textit{Phonon mediated pairing} \cite{choi2018strong,wu2018theory,qin2021critical,qin2021critical,abanin2007quantized,codecido2019correlated,ojajarvi2018competition}:--  Optical (acoustic) phonon mediated interactions are outside (inside) the interaction 
parameter space spanned by $\Gamma^{AA}$ and $\Gamma^{AB}$.  Explicit gap equation solutions 
\cite{wu2018theory,qin2021critical}
show that optical phonons can induce superconductivity if the repulsive Coulomb interaction is somehow 
strongly suppressed.  When the acoustic phonon propagator is not screened by electrons, 
it leads to a $\nu$-independent attraction $\Gamma^{AA}=A^s\sim -0.5$meV \cite{qin2021critical} that is too small to explain the superconductivity dome.  
As shown in Fig.~2, since acoustic phonons reduce $A_s$ \cite{Ceae2107874118}, they can nevertheless play a role in enhancing $T_c$.

The pseudospin parmagnon is able to account for many aspects of the rich phenomenology of MATBG,
 including the mismatch between conditions that favor superconductivity and the anomalous Hall effect,
 the influence of enhanced remote screening, and particle-hole asymmetry.  Our proposal 
 can be tested by systematically studying how encapsulating hBN alignment and remote screening influence superconductivity.
It is interesting to contrast superconductivity in MATBG with the recently 
discovered  \cite{zhou2021superconductivity} superconductivity in ABC trilayer graphene,
which also has spin, valley, and sublattice pseudospins and
has the advantage of simpler underlying electronic structure.
In both cases, pairing is likely to occur between electrons in opposite valleys because the valley-projected 
bands are not time-reversal invariant.
In ABC trilayers superconductivity appears to be enhanced by proximity to a transition that occurs 
between unbroken-symmetry and partially flavor polarized states.  
The difference between the two cases may lie in the underlying physics that controls sublattice
polarization instabilities.

\textit{Acknowledgement:--} We acknowledge informative conversations with Youngjoon Choi, Andrea Young and Haoxin Zhou. 
This work was supported by the U.S. Department of Energy, Office of Science, Basic Energy Sciences, under Award DE-SC0022106.

\bibliographystyle{ieeetr}
\bibliography{reference}

%%%%%%%%%% Merge with supplemental materials %%%%%%%%%%
\newpage
\widetext
\begin{center}
\textbf{\large Supplementary Material: Pseudospin Paramagnons and the Superconducting Dome in Magic Angle Twisted Bilayer Graphene}
\end{center}
%%%%%%%%%% Merge with supplemental materials %%%%%%%%%%
%%%%%%%%%% Prefix a "S" to all equations, figures, tables and reset the counter %%%%%%%%%%
\setcounter{equation}{0}
\setcounter{figure}{0}
\setcounter{table}{0}
\setcounter{page}{1}
\makeatletter
\renewcommand{\theequation}{S\arabic{equation}}
\renewcommand{\thefigure}{S\arabic{figure}}

\section{Non-interacting susceptibilitites in various pseudsopsin channels}
In the main text, we explain the superconducting dome in magic angle twisted bilayer graphene (MATBG) by considering a
sublattice-dependent interaction potential that depends sensitively on the moir\'e band filling factor $\nu$. We focus on $-2<\nu<-3$. We show that MATBG is at the boarder of sublattice polarization over a wide range of $\nu$ and this generates an attractive intra-sublattice interaction between spin-polarized charge carriers in opposite valley and leads to superconductivity. This superconducting channel is suppressed as $\nu\rightarrow -3$ because the valley polarizibility increases and as $\nu \rightarrow-2$ because the density of states at the Fermi level diminish.

In this section, we use a simple bandstructure model to calculate the non-interacting susceptibilities in various pseudospin channels to support our explanation to the superconducting dome.
The bandstructure is obtained from the Bistritzer-MacDonald continuum model at twist angle $\theta=1.15^\circ$ parametrized by interlayer tunneling parameters $t_{AB}=110$meV and $t_{AA}=77$meV for AB and AA bilayer stacking, respectively. We also include a nonlocal momentum-dependent correction to the interlayer tunneling, $dt_{AA}/dk=dt_{AB}/dk=-0.1$ \cite{qin2021inplane}.
Because there is no consensus on the Fermiology in MATBG (e.g.~the location of the Fermi-surface after reset transition at $\nu=-2$), we will only make conclusions that are not sensitive to the details of the bandstructure.

%In the main text, our explanation to the superconducting dome in magic angle twisted bilayer graphene is that the proximity of the system to sublattice polarization instability over a wide range of filling factor $\nu$ generates an attractive intra-sublattice interaction between spin-polarized charge carriers and triggers the valley-singlet superconductivity until $\nu$ approaches $-3$ where the system becomes closer to valley polarization instability or $\nu$ approaches $-2$ where the density of states diminish. 
%In order to quantify how close the system is to instabilities in various pseudospin channels, we evaluate bare particle-hole susceptibilities for the majority-spin electrons in a spin-polarized valley degenerate twisted bilayer graphene at twist angle $\theta=1.15^\circ$ using the non-interacting Bistritzer-MacDonald(BM) model with local intra-and inter-sublattice tunneling $w_1=110$meV, $w_0=77$meV and the non-local tunneling parameter $\eta=0.1$ (See \cite{qin2021inplane} for details of the single-particle Hamiltonian). We define a susceptibility in the particle-hole channel $\mathcal{O}$ as follows:
%

\begin{figure}
    \centering
    \includegraphics[width=1\columnwidth]{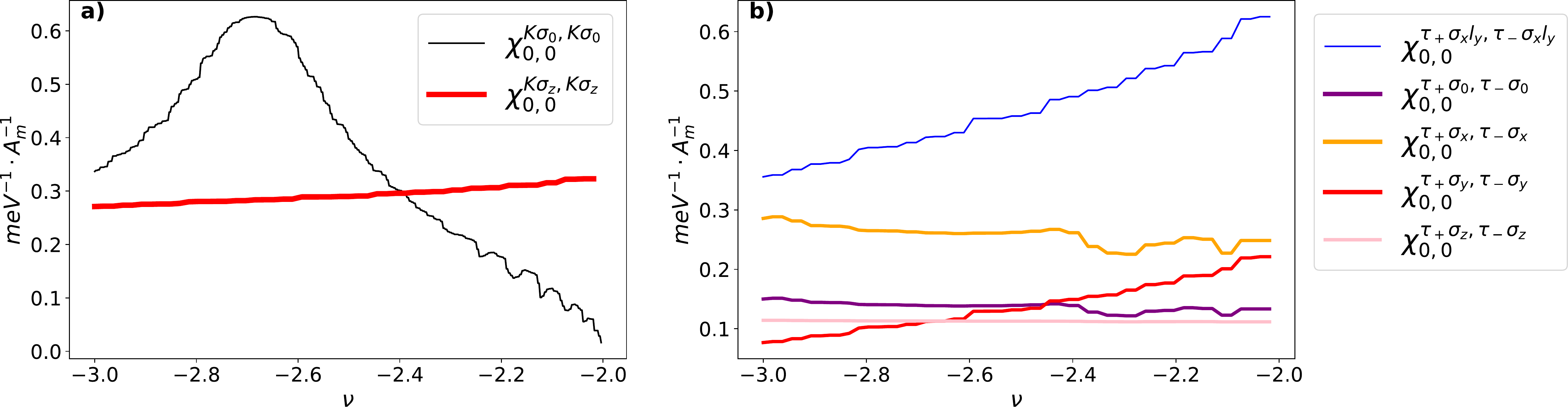}
    \caption{ The static uniform susceptibilities of twisted bilayer graphene in a) pseudospin density channels b) transverse valley channels. $\tau,\sigma,l$ represent valley, sublattice and layer degrees of freedom, respectively. }
    \label{fig:chi}
\end{figure}

As discussed in the main text, the normal state is spin-polarized so we consider only particle-hole excitations in the majority-spin projected Hilbert space. The pseudospin susceptibility of a particle-hole channel $\mathcal{O}$ is defined as follows:
\begin{align}
    &\chi_{\vec{q},\omega}^{\mathcal{O},\mathcal{O}^{\dagger}}=-i\int_{0}^{\infty}dt\ e^{i\omega t}\left\langle\frac{1}{A}\left[\sum_{\vec{p},\alpha\beta}c_{\vec{p},\alpha}^{\dagger}\mathcal{O}_{\alpha\beta}c_{\vec{p}+\vec{q},\beta}\ ,\sum_{\vec{p}',\alpha'\beta'}c_{\vec{p}'+\vec{q},\alpha'}^{\dagger}\mathcal{O}_{\alpha'\beta'}^{\dagger}c_{\vec{p}',\beta'}\right]\right\rangle \label{eq_chipw}\\
    &\qquad\ \ =\frac{1}{A}\sum_{\vec{k}\in\text{B.Z.},\tau m,\tau'n}\frac{n_{F}(\epsilon_{\tau m\vec{k}+\vec{q}})-n_{F}(\epsilon_{\tau' n\vec{k}})}{\omega+\epsilon_{\tau' n\vec{k}}-\epsilon_{\tau m\vec{k}+\vec{q}}+i\delta}|\mathcal{M}_{\tau'n\vec{k},\tau m\vec{k}+\vec{q}}|^2  \label{eq_chi}\\
    &\mathcal{M}_{\tau'n\vec{k},\tau m\vec{k}+\vec{q}}=\sum_{\vec{G}, \alpha\beta}\psi_{n\alpha}^{*}(\vec{k}+\vec{G})\mathcal{O}_{\alpha\beta}\psi_{m\beta}(\vec{k}+\vec{q}+\vec{G}),\qquad \alpha=\tau'\sigma'l'(\beta=\tau\sigma l)\label{eq_me}
\end{align}
% \begin{align}
%     &\chi_{\vec{G},\omega}^{\mathcal{O},\mathcal{O}^{\dagger}}=-i\int_{0}^{\infty}dt\ e^{i\omega t}\left\langle\frac{1}{A}\left[\sum_{\vec{G}_{1,2},\alpha\beta}c_{\vec{k}+\vec{G}_{1},\alpha}^{\dagger}\mathcal{O}_{\alpha\beta}c_{\vec{k}+\vec{G}+\vec{G}_{2},\beta},\sum_{\vec{G}_{1,2}',\alpha'\beta'}c_{\vec{k}+\vec{G}+\vec{G}_{1}',\alpha'}^{\dagger}\mathcal{O}_{\alpha'\beta'}c_{\vec{k}+\vec{G}_{2}',\beta'}\right]\right\rangle \\
%     &\qquad\ \ =\frac{1}{A}\sum_{\vec{k},\tau m,\tau'n}\frac{n_{F}(\epsilon_{\tau m\vec{k}+\vec{G}})-n_{F}(\epsilon_{\tau' n\vec{k}})}{\omega+\epsilon_{\tau' n\vec{k}}-\epsilon_{\tau m\vec{k}+\vec{G}}+i\delta}|\mathcal{M}_{\tau'n\vec{k},\tau m\vec{k}+\vec{G}}|^2  \label{eq_chi}\\
%     &\mathcal{M}_{\tau'n\vec{k},\tau m\vec{k}+\vec{G}}=\sum_{\vec{G}_{1,2}, \alpha\beta}\psi_{\tau'n,\alpha}^{*}(\vec{k}+\vec{G}_{1})\mathcal{O}_{\alpha\beta}\psi_{\tau m,\beta}(\vec{k}+\vec{G}+\vec{G}_{2}),\qquad \alpha=\tau'\sigma'\mu'(\beta=\tau\sigma\mu)\label{eq_me}
% \end{align}
Here $\alpha,\beta$ labels sublattice $\sigma$, valley $\tau$ and layer $l$ degree of freedoms.
In Eq.~\eqref{eq_chi}, the basis is transformed from the plane-wave basis to the single-particle band basis. $|\tau n\vec{k}\rangle=\sum_{\vec{G}\sigma l}\psi_{n\tau\sigma\l}(\vec{k}+\vec{G})|\tau\sigma l,\vec{k+\vec{G}}\rangle$ is a state in band $n$ with energy $\epsilon_{n\vec{k}}$, where $\vec{k}$ is restricted to the first Brillouin zone (chosen to be centered around the Moir\'e $m$ point) and the Bloch wave functions in two valleys are related by $\psi_{n+\sigma l}(\vec{k}+\vec{G})=\psi_{n-\sigma l}^{*}(-\vec{k}-\vec{G})$. $n_{F}$ is the Fermi-Dirac distribution and we set temperature $T=0.44$K. We find that most susceptibilities except for$\chi^{K\sigma_0,K\sigma_0}$ and $\chi^{+\sigma_y,-\sigma_y}$ receive a large contribution from the interband transitions and gradually increases with the ultraviolet energy cutoff \cite{macdonald2012pseudospin}. We truncate to 508 bands per valley (corresponding to energy cutoff $\sim 2.5eV$) and the results are summarized in Fig.\ref{fig:chi}.

% \begin{figure}
%     \centering
%     \includegraphics[width=0.6\columnwidth]{susceptibility.pdf}
%     \caption{ Bare susceptibility in various channels. 
%     }
%     \label{fig:chi}
% \end{figure}

Fig.~\ref{fig:chi}a) illustrates the competition between sublattice even ($\chi_{\vec{q}=0,\omega=0}^{K\sigma_{0},K\sigma_{0}}$) and odd ($\chi_{\vec{q}=0,\omega=0}^{K\sigma_{z},K\sigma_{z}}$) susceptibilities as a function $\nu$. The former (at zero temperature) is equal to the Fermi surface density of states in a single valley $N_F$ and therefore, it depends sensitively on $\nu$.
It vanishes when the Fermi level approaches the Dirac points (which is at $\nu=-2$ in our bandstructure model) and increases rapidly as the hole densities increases. 
In contrast, $\chi_{\vec{q}=0,\omega=0}^{K\sigma_{z},K\sigma_{z}}$ shows very weak $\nu$ dependence because the $q=0$ susceptibility does not have any intra-band contribution when the band has a $\mathcal{C}_2T$ symmetry where $\mathcal{C}_2$ and $T$ are inversion and spinless time-reversal symmetry. As a result, it must be larger than $N_F$ as $\nu\rightarrow-2$. This shows that the sublattice polarizability enhanced by a factor of $[1-(u_0+u_\perp)\chi_{\vec{q}=0,\omega=0}^{K\sigma_{z},K\sigma_{z}}]^{-1}$
is typically stronger than the enhanced valley polarizability 
$N_F/[1-(u_0-u_\perp)N_F]$ except for filling factor close to $\nu=-3$. Recall $u_\perp>0$.
%DOS gradually increases as we hole-dope the system from $\nu=-2$ and become particularly large near $\nu=-3$. In contrast, $\chi_{\vec{q}=0,\omega=0}^{K\sigma_z,K\sigma_z}$ slowly increases with $\nu$ and is largest and surpasses DOS near $\nu=-2$. Note that Coulomb interaction neglected in this calculation tends to broaden the bandwidth of flat bands and significantly lowers DOS, but the sublattice odd susceptibility is expected to be less sensitive to the change of the bandwidth as one can check from the non-interacting BM model by increasing the twist angle and thus the bandwidth, so it is likely that in reality there exists a wider range of filling factors where $\chi_{\vec{q},\omega=0}^{K\sigma_{0},K\sigma_{0}}<\chi_{\vec{q},\omega=0}^{K\sigma_{z},K\sigma_{z}}$ and the system is closer to the sublattice polarization instability than the valley polarization instability.

Fig.~\ref{fig:chi}b) shows the susceptibility associated with neutral excitations where the particle and hole are drawn from opposite valley. There are four layer-diagonal, sublattice dependent susceptibility denoted by $\chi^{{\tau_+}\sigma_i,\tau_-\sigma_i}, i=x,y,z,0$. The mixing between different channels are not considered, a justified approximation in the limit $\vec{q}=\omega=0$ due to the $D_6$ and spinless time reversal symmetry of twisted bilayer graphene. As an example, we demonstrate that the sublattice even and odd intervalley susceptibilities decouple: The $C_{3}$ symmetry acts as exp$(-i\frac{2\pi}{3}\tau_z\sigma_z)$ in the valley-sublattice pseudospin space. It commutes with $\tau_{\pm}\sigma_{x,y}$ but rotate $\tau_{\pm}\sigma_{0,z}$ by a phase factor exp$(\pm i\frac{2\pi}{3})$. Therefore, $\chi_{\vec{q},\omega}^{\tau_+\sigma_{0,z},\tau_-\sigma_{x,y}}=$ exp$(\pm i\frac{2\pi}{3})\chi_{R_{2\pi/3}\vec{q},\omega}^{\tau_+\sigma_{0,z},\tau_-\sigma_{x,y}}$ so it must vanish at $\vec{q}=0$. Since intervalley sublattice even channels do not mix with sublattice odd ones and Fig.~\ref{fig:chi}b) shows that the former have small susceptibilities, they can be neglected in the effective interaction in the main text.
When the particle and hole drawn from opposite valleys are also from opposite layers, the susceptibility is significantly larger than those that are layer-diagonal as shown in Fig.~\ref{fig:chi}b).
The origin of the enhanced susceptibility in this channel is due to a (small-twist-angle) emergent $C_2P=\tau_{x}\sigma_{x}l_{y}$ symmetry in twisted bilayer graphene discovered in Ref.~\cite{lian2021iv}. Because of $C_2P$, $\langle K, -n,\vec{k}|\tau_{+}\sigma_x l_y|K', n,\vec{k}\rangle\approx 1$ where $P$ is a unitary particle-hole symmetry \cite{lian2021iv} and $\pm n$ bands are $C_2P$ partners. The large matrix elements between the $C_2P$ partners result in a large $\chi^{+\sigma_x l_y,-\sigma_x l_y}$ according to Eq.~\ref{eq_me}. Note that this susceptibility dominates other pseudospin particle-hole channels when the Fermi level approached the Dirac points and might lead to an intervalley exciton condensation.
The resulting valley XY-like order parameter $\tau^{+}\sigma_x l_y$ preserves a spinless time reversal symmetry $\mathcal{T}=\tau_y\mathcal{K}$($\mathcal{T}^2=-1$), and hence is consistent with the previously proposed Kramers-intervalley coherent order at $\nu=-2$ \cite{bultinck2020ground,lian2021iv}. 
%\textcolor{blue}{Note the sublattice and layer off-diagonal particle-hole fluctuation near the valley-XY instability generates a strong inter-sublattice repulsion via the Berk-Schrieffer diagram, but it does not mix into the intra-sublattice and intra-layer attractive interaction. We therefore believe that it will not significantly suppress the intra-sublattice pairing in our theory, analogous to the effect of the inter-sublattice intra-layer repulsion $\Gamma^{AB}$ shown in Fig.2 in the main text. Nevertheless, to avoid the uncertainty on the sign of the inter-layer interaction, we only consider the intra-layer interaction in the model discussed in the main text and in the following sections.}
%
%
Note the tendency towards valley-XY order generates a strong valley-singlet repulsion via the Berk-Schrieffer diagram, as discussed in the main text. However, because this repulsion is also off-diagonal in sublattice (like $\Gamma^{AB}$) and the $C_2$ and $C_3$ symmetry decouple interaction according to their sublattice parity in the linearized gap equation (as shown in the next section), superconductivity prevails as long as intra-sublattice is attractive ($\Gamma^{AA}<0$). 
In the main text and in what follows, we neglect valley-singlet repulsion generated by all the valley-XY fluctuations since the channels that modify intra-sublattice attraction is very small. In this case, suppression of valley-singlet superconductivity close to $\nu=-2$ will only require a mundane explanation -- a reduced density of states.

We would also like to point out that in the definition of the susceptibility in Eq.~\eqref{eq_chi} the particle-hole momentum $\vec{q}$ is, rigorously speaking, conserved only up to moir\'e reciprocal vectors $\vec{G}$ and therefore the bare susceptibility $\chi_{\vec{q},\omega}$ used in the generalized RPA formula in the main text should be understood as a matrix in the momentum space, $\hat{\chi}_{\vec{q},\omega}(\vec{G},\vec{G}')$. However, for the purpose of comparing the relative strength of susceptibilities in various channels, it is adequate to focus on the small $\vec{q}$ and $\vec{G}=\vec{G}'=0$.

\section{Linearized Gap equation}
In this section, we study the linearized gap equation within the reduced interaction model described by Eqs.~(1) and (2) in the main text. The real-space interaction model is given by the Hamiltonian
\begin{equation}
\begin{aligned} 
\label{eq:H_reds}
H_{\text{red}} & = \int d^2\vec{r}   \sum_{l,l' \tau \sigma \sigma'} \Gamma^{\sigma\sigma'} \delta_{l,l'} \delta_{\tau,-\tau'}\, \psi^{\dagger}_{l \tau \sigma}(\vec{r}) \psi^{\dagger}_{l' \tau' \sigma' }(\vec{r})   \psi_{l' \tau' \sigma' }(\vec{r})  \psi_{l 
\tau  \sigma}(\vec{r}),
\end{aligned}
\end{equation}
where $l,l'$ label layer, $\tau,\tau'$ label valley and $\sigma,\sigma'$ denote sublattice. The interaction matrix $\Gamma^{\sigma\sigma'} = \Gamma^{AA} \delta_{\sigma,\sigma'} +\Gamma^{AB}\delta_{\sigma,-\sigma'} $. 
This leads to the following linearized gap equation
\begin{equation}
\Delta^{l}_{\sigma \sigma'}(\vec{G})  =  \Gamma^{\sigma\sigma'}  \sum_{l'\sigma_1 \sigma_1'\vec{G}'}    \Pi_{l\sigma \sigma' ,l'\sigma_1 \sigma_1'}  (\vec{G},\vec{G}') \Delta^{l'}_{\sigma_1 \sigma_1'}(\vec{G}'),
\end{equation}
where $\vec{G}$ and $\vec{G}'$ are moir\'e reciprocal lattice vectors. 
The layer and sublattice-dependent pair susceptibility
\begin{equation}
\Pi_{l\sigma \sigma' ,l'\sigma_1 \sigma_1'}  (\vec{G},\vec{G}') =  \sum_{m m' \vec{k}'}  \frac{n_F(\epsilon_{m\vec{k}'})-n_F(-\epsilon_{m'\vec{k}'}) }{\epsilon_{m\vec{k}'}+\epsilon_{m'\vec{k}'}} \langle m' \vec{k}'| e^{i\vec{G}\cdot \vec{r}}  | m \vec{k}' \rangle_{l\sigma'\sigma} \langle m' \vec{k}'| e^{i\vec{G}'\cdot \vec{r}} |m \vec{k}'\rangle_{l'\sigma_1'\sigma_1}^*, 
\label{eq:scsus}
\end{equation}
where $m$ and $m'$ are band indices. For the above contact-interaction model, the interaction matrix is independent of moir\'e momentum $\vec{k}$, suggesting the superconducting pairing potential exhibits moir\'e periodicity in the real space.  The critical temperature is obtained by equating the largest eigenvalue of matrix $ -\Gamma \cdot \Pi $ to 1. We numerically diagonalized this matrix as a function of the input $(\Gamma^{AA},\Gamma^{AB})$ and this leads to the result shown in Fig.~1 of the main text. The results we obtained are not qualititively modified by the factor
$\delta_{l,l'}$ we used to simplify Eq.~\eqref{eq:H_reds}.

For MATBG, the largest contribution to the pair susceptibility come from the flat conduction and valence bands and the Bloch wavefunctions from the these bands are localized at small moir\'e $G$-vectors (i.e.~they vary slowly in the moir\'e unit cell).
Therefore, although the eigenvector of the Eq.~5 has a very large dimension, most of the properties of the superconducting order parameter can be understood in the subspace of $\vec{G} = \vec{G}' = 0$. In this subspace, the pairing amplitude of a spin-polarized valley-singlet superconductor can be conveniently described by the following $2\times2$ matrix,
\begin{equation}
F_{\sigma \sigma'}(\vec{k})=\langle c_{K\sigma \uparrow}(\vec{k})c_{K'\sigma'\uparrow}(-\vec{k})\rangle,\label{eq_pairingamp}
\end{equation}
where we assume all electrons are from the same layer. The linearized gap equation takes a much simpler form,
\begin{equation}
\Delta_{\sigma \sigma'}(\vec{k})=\sum_{\vec{k}'\sigma_1\sigma_1'}\Gamma_{\vec{k}\vec{k}'}^{\sigma \sigma'} \Pi_{\sigma \sigma',\sigma_1\sigma_1'}(\vec{k}')\Delta_{\sigma_1\sigma_1'}(\vec{k}').
\label{eq_gap}
\end{equation}
We reinserted the momentum $\vec{k},\vec{k}'$ dependence above for the clarity of analysis below.
This sublattice dependent gap equation for MATBG in the $\vec{G}=\vec{G}'=0$ subspace is reminiscent to the spin-dependent gap equation in $^3$He.  However, unlike $^3$He, the sublattice pseudospin conservation is broken in MATBG by the non-interaction single-particle Hamiltonian. 
Because the order parameter matrix must be anti-symmetric under transposition, \textit{i.e.} $\Delta(\vec{p}) = -\Delta^{\text{T}}(-\vec{p})$, where $\vec{p}=\vec{K}+\vec{k}$, and the parity operator $\mathcal{C}_2$ flips sublattice polarization $\sigma \rightarrow -\sigma$ and momentum $\vec{p}\rightarrow-\vec{p}$, we have
$\mathcal{C}_{2}\Delta_{\sigma,\sigma'}(\vec{p})\rightarrow
\Delta_{-\sigma ,-\sigma'}(-\vec{p})=-\Delta_{-\sigma' ,-\sigma}(\vec{p})$.
Here we note that, unlike spin in $^3$He, the pseudospin is flipped by the parity operator and this means the even-parity is a singlet while the odd-parity is a triplet,
\begin{align}
\Delta(\vec{p}) & =g_{0}(\vec{p})\sigma^{z}+(\vec{d}(\vec{p})\cdot\vec{\sigma})\sigma^{z}=\begin{pmatrix}g_{0}+d^{z} & d^{+}\\
d^{-} & -g_{0}+d^{z}
\end{pmatrix},
\end{align}
where $g_{0}(-\vec{p})=g_{0}(\vec{p})$ and $\vec{d}(-\vec{p})=-\vec{d}(\vec{p})$.  
The normal state $\mathcal{C}_{2}$ symmetry then decouples the (sublattice-pseudospin) singlet-triplet channel of the gap equation:
\begin{equation} 
g_{0}(\vec{k})=\sum_{\vec{k}'}\Gamma_{\vec{k}\vec{k}'}^{AA}\Pi^{zz}(\vec{k}')g_{0}(\vec{k}'),\label{eq_gapeven}
\end{equation}
\begin{equation}  
\begin{pmatrix} 
d^{z}\\
d^{+}\\
d^{-}
\end{pmatrix}_{\vec{k}}=
\sum_{\vec{k}'}\hat{\Gamma}_{\vec{k}\vec{k}'}
\begin{pmatrix}
\Pi^{00} & \Pi^{0-} & \Pi^{0-}\\
\Pi^{0-} & \Pi^{++} & \Pi^{+-}\\
\Pi^{0-} & \Pi^{+-} & \Pi^{++}
\end{pmatrix}_{\vec{k}'}
\begin{pmatrix}
d^{z}\\
d^{+}\\
d^{-}
\end{pmatrix}_{\vec{k}'}\label{eq_gapfourier}
\end{equation}
where $\hat{\Gamma}_{\vec{k}\vec{k}'}=\text{diag}\{\Gamma_{\vec{k}\vec{k}'}^{AA},\ 2\Gamma_{\vec{k}\vec{k}'}^{AB},\ 2\Gamma_{\vec{k}\vec{k}'}^{AB}\}$, and  
\begin{equation}
\Pi^{ij}(\vec{k})=\frac{1}{2}\sum_{\sigma \sigma',\sigma_1\sigma_1'}(\sigma^{i})_{\sigma \sigma'} (\sigma^{j})_{\sigma_1\sigma_1'} \Pi_{\sigma \sigma',\sigma_1\sigma_1'}(\vec{k}),\ \ i,j=0,z,\pm.
\label{eq_Green2}
\end{equation}
Since superconductivity is an instability of the Fermi surface, it is unlikely to happen in the even-parity channel (i.e.~$g_{0}= 0$) due to the following reasons. First, the intra-band contribution to even-parity pairing susceptibility $\Pi^{zz}$ vanishes since the sublattice polarization is not established in the normal state, indicating $\langle m\vec{k}|\sigma^z| m\vec{k}\rangle = 0$ in Eq.~(\ref{eq:scsus}). Secondly, the inter-band contribution to $\Pi^{zz}$ is negligible because the critical temperature $T_c \ll$ band separation. When $g_{0}= 0$, $\vec{d}$ alone cannot give rise to finite sublattice polarization, so  $\langle\sigma_z\rangle=0$ in the superconducting state and it remains nearly sublattice polarized for $T<T_c$.

%\textcolor{blue}{CL: Wei, let me know if you agree.}
In the odd-parity triplet subspace, when $\Gamma^{AA}$ and $\Gamma^{AB}$ are momentum independent and $\Gamma^{AB}\gg - \Gamma^{AA}>0$, 
the gap equation admits a simple self-consistent solution with $d^{\pm}(\vec{k})=0$  and $d^z(\vec{k})=d^z$ (i.e.~independent of $k$). This paramaterization  simplifies the $3\times3$ matrix equation (Eq.~\eqref{eq_gapfourier}) to the following equation:
\begin{equation}
    \Gamma^{AA}\sum_{\vec{k}} \Pi^{00}(\vec{k},T_c)=-1\;\;,\;\;\Gamma^{AA}<0.
\end{equation}
This simplification occurs because $\sum_{\vec{k}} \Pi^{0-}(\vec{k}) =0$ which follows from $\mathcal{C}_{3}$ rotation symmetry: $\Pi^{0-}(\mathcal{C}_{3}\vec{k}) = e^{- 4\pi i/3}\Pi^{0-}(\vec{k})$. Clearly, $T_c$ obtained from this simplified gap equation is independent of $\Gamma^{AB}$. This means when the two members of the Cooper pairs are from the same sublattice, they can avoid the strong inter-sublattice repulsion $\Gamma^{AB}$ if the pairing amplitude is uniform on the Fermi surface. 
%In the numerics, when $\Gamma_{AB}$ is 20 times stronger than $\Gamma_{AA}$, the $T_c$  started to decrease and this stems from larger moir\'e reciprocal lattice vectors that are neglected in the above analysis. 

\begin{figure}
    \centering
    \includegraphics[width=0.5\columnwidth]{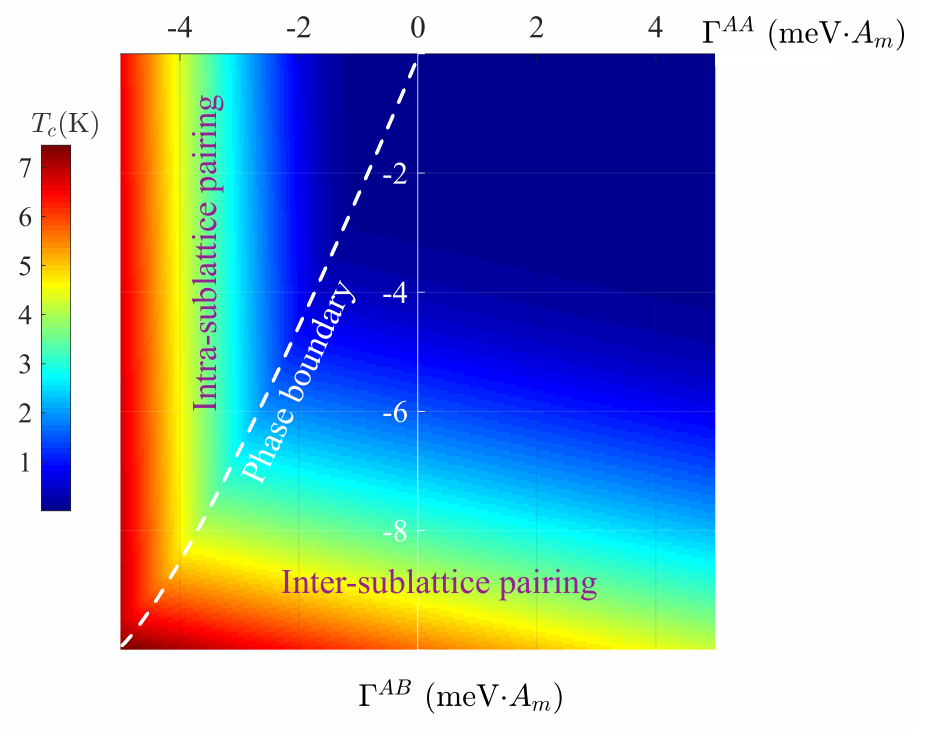}
    \caption{ Color plot of critical temperature $T_c$ as function of $\Gamma_{AA}$ and $\Gamma_{AB} \le 0$, where $A_m$ denotes the area of moir\'e supercell. The dashed curve highlights the phase boundary between the intra- and inter-sublattice pairing states. These numerical results are obtained by using a continuum model for MATBG with $t_{AA}/t_{AB}=0.7$ and filling factor $\nu=-2.4$.  
    }
    \label{fig:figures2}
\end{figure}

While we do not have compelling arguments to think that MATBG will be located in the parameter space where $\Gamma^{AB}$ is attractive, it nevertheless lead to superconductivity, as shown in Fig.~\ref{fig:figures2}. Let us discuss the structure of the superconducting order parameter in this parameter space without dwelling with the microscopic origin of an attractive $\Gamma^{AB}$.
In the parameter space where $\Gamma^{AA}\gg -\Gamma^{AB}>0
$, we can write down two degenerate self-consistent solutions $(d^{z}(\vec{k}),d^{+}(\vec{k}),d^{-}(\vec{k}))= (0,d^+,0)$ 
and $(d^{z}(\vec{k}),d^{+}(\vec{k}),d^{-}(\vec{k}))= (0,0,d^-)$ to reduce the $3\times3$ matrix equation to a single gap equation which is independent of $\Gamma^{AA}$:
\begin{equation}
    \Gamma^{AB}\sum_{\vec{k}} \Pi^{++}(\vec{k},T_c)=
     \Gamma^{AB}\sum_{\vec{k}} \Pi^{--}(\vec{k},T_c)=-1
     \;\;,\;\;\Gamma^{AB}<0\;\;.
\end{equation}
This is again due to the property $\sum_{\vec{k}} \Pi^{+-}(\vec{k}) =0$ which follows from $\mathcal{C}_3$: $\Pi^{+-} (\mathcal{C}_{3}\vec{k})= e^{ 4\pi i/3}\Pi^{+-} (\vec{k})$. The phase boundary marked in Fig.~2 denotes the change of superconducting order parameter from intrasublattice-like $\vec{d}(\vec{k})=d^z$ to intersublattice-like $\vec{d}(\vec{k})=d^+ \; \text{or}\; d^-$.
In this simple analysis, we can always reduce the $3\times3$ linearized gap equation for the $\mathcal{C}_2$ triplet to a scalar equation which only involves either $\Gamma^{AA}$ or $\Gamma^{AB}$. Hence, intra-sublattice Cooper pairs do not `see' $\Gamma^{AB}$ and the 
inter-sublattice Cooper pairs  do not `see' $\Gamma^{AA}$.
When we account for non-local correction from larger moir\'e reciprocal vector, such perfect decoupling of $\Gamma^{AA}$ and $\Gamma^{AB}$ is no longer exact and $T_c$ can indeed be suppressed by the repulsion. For reasons that are unclear to us, such suppression occurs in a very asymmetrical way. The suppression of intra-sublattice pairing ($\Gamma^{AA}<0$) via repulsive $\Gamma^{AB}$ only becomes significant when the repulsion exceeds the attraction by a factor of $10$, $\Gamma^{AB}\sim -10\Gamma^{AA}>0$. 
On the other hand, the suppression of inter-sublattice pairing ($\Gamma_{AB}<0$) from repulsive $\Gamma^{AA}$ is very effective -- when the repulsion $\Gamma^{AA}$ is roughly the same as the attraction $\Gamma^{AB}$, the $T_c$ is already negligibly small.

\section{Valley-triplet, spin-singlet Pairing and Berk-Schrieffer diagrams}

In the maintext we assume the two members of a Cooper pair are from opposite valleys because they are related by a (spinless) time-reversal symmetry. When they are drawn from the same valley, the band-energy is usually pair breaking but it is nevertheless interesting to carry out the paramagnon analysis in this valley-triplet channel. When the superconducting order parameter is uniform, valley-triplet has to be a spin-singlet so the four external particle-lines of the particle-particle irreducible interaction $\Gamma$ shown in Fig.~3a of the main text are given by $1,2,3,4=K \uparrow ,K\downarrow,K\downarrow,K\uparrow$. 
Since pairing occurs in the valley-projected Hilbert space, 
the diagrammatics is similar to the original Berk-Schrieffer analysis \cite{berk1966effect} with an extra sublattice degree of freedom.
In particular, the pseudospin-density fluctuation contribution to $A^a$ is
%Since $u_\perp$ does not appear in the analysis, the usual Coulomb interaction $u_0$ simply enhances spin-polarizability and hence suppresses spin-singlet SC. 
%
\begin{align}
    A^a =& \frac{1}{2} \left( T^{\sigma_zs_0}-T^{\sigma_zs_z} \right)\nonumber \\
    =&\frac{1}{4}
    \left(\frac{-u_0}{1-u_0\chi^{\sigma_zs_0,\sigma_zs_0}} +
    \frac{u_0}{1-u_0\chi^{\sigma_zs_z,\sigma_zs_z}} \right)=0.
\end{align}
The last equation is due to $\chi^{\sigma_z s_z,\sigma_z s_z}=\chi^{\sigma_zs_0,\sigma_z s_0}$ and it happens because the valley-projected space is fully spin-polarized. As a result, sublattice fluctuations do not generate attractive interaction in the spin-singlet valley-triplet channel.

Let us now elaborate on the discuss of the
Berk-Schrieffer (BS) transverse fluctuation contributions 
indicated in Fig.~3 of the main text. The particle-hole channels in the BS diagrams 
decouple into four channels $\tau_{+}\sigma_{0,x,y,z}$ in the long wavelength limit and they change
\begin{equation}
    A^{s,a}\rightarrow A^{s,a} \mp \frac{1}{2} \sum_{i=x,y}T^{\tau_+\sigma_i} -\frac{1}{2} \sum_{\delta=0,z}T^{\tau_+\sigma_\delta},
\end{equation}
where the intervalley intrasublattice interaction  $T^{\tau_+\sigma_\delta}$ is suppressed by small matrix-elements at the vertex because they mix states from opposite Chern band while the inter-valley component is given by the following:
\begin{align} \label{eq:T_xy_z}
   T^{\tau_{+}\sigma_{i}}&=-\frac{u_{0}^{2}\chi_{\vec{k}+\vec{k}',\omega}^{\tau_{+}\sigma_{i},\tau_{-}\sigma_{i}}}
   {1-u_{0}\chi_{\vec{k}+\vec{k}',\omega}^{\tau_{+}\sigma_{i},\tau_{-}\sigma_{i} }} \; ,\; i=x,y.
\end{align}
Here the susceptibility $\chi$ is shown in Fig.~1b.
While this leads to a repulsive $\Gamma^{AB}$, it does not modify $\Gamma^{AA}$ hence superconductivity prevails as shown in the maintext (Fig.~1). If the interaction is mainly enhanced at zero paramagnon momentum $q=0$, then valley-Ising paramagnon ($T^{\tau^{z}\sigma_0}$) with $\vec{q}=\vec{k}-\vec{k}'$ scatters Cooper pairs locally around the Fermi surface  while the valley-XY paramagnon  ($T^{\tau_{+}\sigma_{x,y}}$) with $\vec{q}=\vec{k}+\vec{k}'$ scatters Cooper pairs across the Fermi surface, the latter is similar to the short-range repulsive interaction generated by antiferromagnetic paramagnon in cuprates \cite{scalapino1999superconductivity}. We summarize important information about different pseudospin paramagnons in Table.~1. 

\begin{table}
\caption{A summary of important properties for the charge fluctuation and 7 pseudospin paramagnons.}
\begin{tabular}{|c|c|c|c|c|c|c|}
\hline 
$T$ & $\tau_{0}\sigma_{0}$ & $\tau_{0}\sigma_{z}$ & $\tau_{z}\sigma_{z}$ & $\tau_{z}\sigma_{0}$ & $\tau_{+}\sigma_{x,y}$ & $\tau_{+}\sigma_{0,z}$\tabularnewline
\hline 
\hline 
Types of fluctuation & \multicolumn{4}{c|}{Charge and Pseudospin Density} & \multicolumn{2}{c|}{Berk-Schrieffer }\tabularnewline
\hline 
Dominant scattering vertex & Direct & \multicolumn{5}{c|}{Exchange}\tabularnewline
\hline 
Exchange enchancement & -- & \multicolumn{2}{c|}{$-4<\nu<4$} & $|\nu|-2\rightarrow1$ & $|\nu|-2\rightarrow0$ & -- \tabularnewline
\hline 
Types of Order & -- & Sublattice-Polarized & Orbital-Moment & Valley-Ising & Valley-XY (IVC) & --\tabularnewline
\hline 
\end{tabular}
\end{table}

\end{document}